\documentclass[12pt]{article}
\usepackage{amsmath,amssymb,bm,graphicx,epsfig,subfigure}
\usepackage{lineno,booktabs,url}
\include{epsf} 
\newcommand\as{\alpha_S}

\def\beqn{\begin{eqnarray}} 
\def\eeqn{\end{eqnarray}} 
\def\beq{\begin{equation}} 
\def\eeq{\end{equation}}

\begin{document}
\begin{titlepage}
\begin{flushright}
IFIC/21-51\\
FTUV-21-1129.5289
\end{flushright}

\renewcommand{\thefootnote}{\fnsymbol{footnote}}
\vspace*{1cm}

\begin{center}
{\Large \bf 
  Fiducial perturbative power corrections \\  [0.1cm]
 within the $\bf q_T$ subtraction formalism
}
\end{center}

\par \vspace{2mm}
\begin{center}
  {\bf Stefano Camarda${}^{(a)}$},  {\bf Leandro Cieri${}^{(b)}$}
  and {\bf Giancarlo Ferrera${}^{(c)}$}\\

\vspace{5mm}

${}^{(a)}$ 
CERN, CH-1211 Geneva, Switzerland\\\vspace{1mm}


${}^{(b)}$ 
Instituto de F\'isica Corpuscular, Universitat de Val\`encia - Consejo Superior de
Investigaciones Cient\'ificas, Parc Cient\'ific, E-46980 Paterna, Valencia, Spain\\\vspace{1mm}

${}^{(c)}$ 
Dipartimento di Fisica, Universit\`a di Milano and\\ INFN, Sezione di Milano,
I-20133 Milan, Italy\\\vspace{1mm}

\end{center}

\vspace{1.5cm}

%
\par \vspace{2mm}
\begin{center} {\large \bf Abstract} \end{center}
\begin{quote}
  \pretolerance 10000
  We consider higher-order QCD corrections to the production of
  high-mass systems in hadron collisions within the transverse-momentum
  ($q_T$) subtraction formalism.
  We present a method to consistently remove the linear power corrections in $q_T$
  which appears when fiducial
  kinematical  cuts 
  are applied on the final state system. 
  We consider explicitly the case of fiducial cross sections for
  Drell-Yan lepton pair production
  at the Large Hadron Collider
  up to next-to-next-to-next-to-leading order (N$^3$LO)
  in QCD.
We have implemented our method within the  {\ttfamily DYTurbo} numerical program
and we have obtained perturbative predictions which are in agreement at
the permille
level with those obtained with local subtraction formalisms up to the next-to-next-to-leading order (NNLO). At the N$^3$LO we are able to provide predictions for fiducial cross sections with numerical accuracy at the permille level.


\end{quote}

\vspace*{\fill}
\vspace*{2.5cm}

\begin{flushleft}
November 2021
\end{flushleft}
\end{titlepage}

\setcounter{footnote}{1}
\renewcommand{\thefootnote}{\fnsymbol{footnote}}

Hard scattering processes at high-energy colliders, such as the Large Hadron Collider (LHC), characterized by large scales of energy ($M$) transferred,  allows us to probe the
dynamics of fundamental interactions at short distances. In this regime, theoretical predictions for cross sections  
can be evaluated with perturbative techniques. 
In particular accurate results require the inclusion of the dominant effects from strong interactions through the calculation
of the higher-order terms in Quantum Chromodynamics (QCD) as a series expansion in the coupling $\alpha_S(M)$. 
In order to match the experimental kinematical cuts on the measured final states, it is essential to obtain predictions for fiducial cross sections
 and corresponding differential distributions.

The computation of higher-order QCD corrections at fully-differential level is complicated by the presence of infrared singularities at intermediate
stage of the calculation which prevents a direct implementation of numerical techniques and enforce the use of an hybrid analytic and numerical approach.
At the next-to-leading order (NLO) general {\itshape subtraction} algorithms, which   exploit the universality properties of soft and collinear emissions in QCD, have been
developed\,\cite{Catani:1996jh,Catani:1996vz,Frixione:1995ms,Frixione:1997np}.
These methods have been successfully implemented in general purpose Monte Carlo programs which
satisfy the needs for the analysis of experimental data.
Beyond the NLO, a widely used extension of the subtraction method  is the
so called transverse-momentum ($q_T$) subtraction
formalism originally proposed in Ref.\,\cite{Catani:2007vq}.
In fact, thanks to its relative simplicity and generality, the method has been successfully
applied to fully differential
QCD calculations for several hard-scattering processes 
at the next-to-next-to-leading order (NNLO) (see Ref.\,\cite{Grazzini:2017mhc} and references therein)
and, more recently, also at the next-to-next-to-next-to-leading order N$^3$LO\,\cite{Cieri:2018oms,Camarda:2021ict}.

In the case of the $q_T$ subtraction formalism, a source of numerical uncertainty which is  particularly difficult to quantify in a robust way is due to the
unphysical power corrections  of the type $\mathcal{O}(q_T^{\rm cut}/M)$, where $q_T^{\rm cut}$ is the  technical parameter necessary to separate the $q_T$ resolved and unresolved parton emissions.
Power corrections ambiguities are particularly severe in the case of fiducial selection cuts which yield an acceptance that has a linear dependence
on $q_T^{\rm cut}$\,\cite{Ebert:2019zkb,Ebert:2020dfc,Alekhin:2021xcu}. 
In principle
the effect of these perturbative power corrections 
can be reduced setting the value of the technical parameter $q_T^{\rm cut}$
sufficiently small. However, very small values of $q_T^{\rm cut}$ require an extremely precise numerical control
of cross sections in the infrared region $q_T\to 0$ which are typically very challenging and
time consuming.

In the case of the production of colourless high mass systems,
sufficiently inclusive cross sections (such as total cross section in absence of fiducial selection cuts)
computed within the $q_T$ subtraction method
have a residual dependence on $q_T^{\rm cut}$ of order
$\mathcal{O}((q_T^{\rm cut}/M)^2)$ originated from the
integration of the corresponding scattering amplitudes
over the final state kinematics in the small $q_T$ region\,\cite{Moult:2016fqy,Cieri:2019tfv}.
In the case of the production of coloured systems the residual dependence on
$q_T^{\rm cut}$ is linear even in absence of fiducial cuts\,\cite{Catani:2014qha,Bonciani:2015sha,Buonocore:2019puv}.
In this case, in order to remove the linear power corrections in $q_T^{\rm cut}$, it is necessary to take care of both
the dependence related to the fiducial  selection cuts and the dependence which is present at the inclusive level.

Linear fiducial power corrections have been connected 
with alternating sign factorial growth of the perturbative expansion   
in Ref.\,\cite{Salam:2021tbm}
and  
modifications of the selection cuts typically used
in experimental analysis have been proposed
in order to eliminate the linear dependence of
the acceptance.
In Ref.\,\cite{Glazov:2020gza}, an experimental procedure has been proposed to remove from cross section measurements
the effect of selection cuts which are responsible for linear fiducial power corrections.
In Ref.\,\cite{Ebert:2020dfc}
it has been shown that linear fiducial power corrections can be consistently
removed through the $q_T$ resummation formalism,
if the $q_T$ recoil due to multi-parton emission
is correctly taken into account.

Fixed-order calculations has a great relevance in precision physics at colliders: they can be computed (in principle)  in a definite and unambiguous way,
and they are an essential ingredient for all-order resummed predictions.
Therefore the goal of having at disposal fixed-order calculations at high numerical accuracy is very relevant,
regardless of the effective {\itshape physical} precision of such predictions.

In this paper we consider standard fixed-order calculations and we discuss a method to remove linear fiducial  power corrections (FPC) within
the $q_T$ subtraction formalism. 
Our method, which is analogous to the one proposed in Ref.\,\cite{Ebert:2020dfc} based on a Lorentz decomposition for
hadronic and leptonic tensors, introduces an additional subtraction exploiting the recoil procedure of Ref.\,\cite{Catani:2015vma}
and it allows us to obtain fiducial fixed-order predictions with a residual uncertainty of the type $\mathcal{O}((q_T^{\rm cut}/M)^2)$ which can be
brought down at sub-permille level. 
Our results turn out to be  crucial in the case of the N$^3$LO extension of the $q_T$ subtraction formalism\,\cite{Cieri:2018oms,Camarda:2021ict} where
it is particularly challenging to obtain precise perturbative predictions for very small values of  $q_T^{\rm cut}$.

We consider explicitly the case of fiducial cross sections for   Drell-Yan lepton pair production   at the Large Hadron Collider
up to the N$^3$LO  in QCD. We have implemented our method within the  {\ttfamily DYTurbo}\,\cite{Camarda:2019zyx} numerical program
and we have obtained perturbative predictions which are in agreement at permille level with those obtained with local subtraction formalism at NLO and NNLO.

We consider the hard-scattering process
\beqn
h_1(p_1) + h_2(p_2) 
\to \sum_i F_i(q_i)+X,
\eeqn
where $F_i$ denotes the (colourless) final states with momenta $q_i$
produced by the colliding hadrons $h_1$ and $h_2$
which we collectively identify as the system $F(q)$, with momentum
$q=\sum_i q_i$, invariant mass $M=\sqrt{q^2}$ and transverse momentum
$q_T$.

We start from the 
the  master formula of the $q_T$ subtraction
formalism for the hadronic cross section
\,\cite{Catani:2007vq}
\beqn
d\sigma^F=d\sigma^{F}_{LO}\otimes \mathcal{H}^F 
+\left[d\sigma^{\rm F+jets}-d\sigma^{\rm CT}\right]\,,
\label{qTsub}
\eeqn
where $\sigma^{F}_{LO}$  is the Born level cross section, 
$\mathcal{H}^F(\alpha_S)$ is the process-dependent hard-collinear function\,\cite{Catani:2012qa,Catani:2013tia} with the following perturbative expansion
\begin{equation}
\label{hexpan}
{\cal H}^F=
1+ \frac{\alpha_S}{\pi} \,{\cal H}^{F\,(1)} 
+ \left(\frac{\alpha_S}{\pi}\right)^2 
\,{\cal H}^{F\,(2)}
+ \left(\frac{\alpha_S}{\pi}\right)^3 
\,{\cal H}^{F\,(3)}+\sum_{n=4}^{\infty}\left(\frac{\alpha_S}{\pi}\right)^n\,{\cal H}^{F\,(n)} \;,
\end{equation}
$d\sigma^{\rm CT}$ is the subtraction counter-term\,\cite{Bozzi:2005wk}
\beqn
d\sigma^{\rm CT} = d\sigma^{F}_{LO} \otimes \Sigma^{F}(q_T/M) d^2{\bf q_T}
\label{qTCT}
\eeqn
and the symbol $\otimes$ stands for convolutions over momentum fractions and sum over flavour indices of the partons.
The second term in the r.h.s.\ of Eq.\,(\ref{qTsub}), $d\sigma^{\rm F+jets}$, is the differential cross section for the
production of $F(q)$ in association with jets and it has to be
evaluated
at the previous perturbative order with respect to $d\sigma^{\rm F}$.
The  subtraction counter-term  $d\sigma^{\rm CT}$ has the same singular behaviour of $d\sigma^{\rm F+jets}$ in the limit $q_T \to 0$
which functional form is known
from the $q_T$ resummation formalism\,\cite{Catani:2000vq,Bozzi:2005wk}.
 
The terms $d\sigma^{\rm F+jets}$ and $d\sigma^{\rm CT}$ in Eq.\,(\ref{qTsub})
are separately divergent due to infrared singularities at $q_T=0$ and a technical parameter
$q_T^{\rm cut}$ has to be introduced.
For $q_T^{\rm cut}>0$ the sum of the terms in the square bracket of Eq.\,(\ref{qTsub}) is IR finite (or, more precisely, integrable over $q_T$)
and it should be evaluated in the limit $q_T^{\rm cut}\to 0$
to obtain the ``exact'' (free form residual $q_T^{\rm cut}$ dependence)
value of the cross section.
However for finite value of $q_T^{\rm cut}$ the cross section in  Eq.\,(\ref{qTsub}) contains power corrections
$\mathcal{O}((q_T^{\rm cut}/M)^p)$, with $p>0$\,\cite{Ebert:2019zkb}. The exact value of the exponent $p$ depends
by the cuts on the final states which define the fiducial cross section
\beqn
\sigma^F_{\rm fid} = \int_{\rm cuts} d\sigma^F\,, 
\label{qTCT2}
\eeqn
we thus have
\beqn
\sigma^F_{\rm fid} =\int_{\rm cuts} d\sigma^{F}_{LO}\otimes \mathcal{H}^F 
+\int_{{\rm cuts}} \left[d\sigma^{\rm F+jets}_{q_T>q_T^{\rm cut}}-d\sigma^{\rm CT}_{ q_T>q_T^{\rm cut}}\right]+\mathcal{O}\left((q_T^{\rm cut}/M)^p\right)\,.
\label{qTsub2}
\eeqn

In Ref.\,\cite{Alekhin:2021xcu} has been shown that, in the case of Drell--Yan lepton pair production, typical cuts on the transverse momenta and rapidities of the final
state particles $F_i(q_i)$ leads to linear power corrections ($p=1$), which
corresponding systematic uncertainty may
spoil the accuracy of fixed-order
calculations within the $q_T$ subtraction formalism.

Clearly 
the effect of perturbative power corrections $\mathcal{O}((q_T^{\rm cut}/M)^p)$
can be reduced setting the value of the technical parameter $q_T^{\rm cut}$
sufficiently small. However, very small values of $q_T^{\rm cut}$
lead to large cancellations among the terms in the square bracket of
the r.h.s.\ of Eq.(\ref{qTsub}), which in turns give rise to larger
numerical integration uncertainties. These cancellations are particularly
challenging at NNLO and N$^3$LO where the precise knowledge of the
fully differential calculations of $F$ in association with jets at NLO and NNLO is respectively
required.
Eventually a trade-off
between  the
systematical and statistical uncertainties of the computation
have to be found   and,
more importantly, a robust systematic uncertainty to the missing
perturbative power corrections has to be computed.
The systematic uncertainty
can be estimated by evaluating the
cross section at different values of $q_T^{\rm cut}$ and
carrying out a $q_T^{\rm cut}\to 0$ extrapolation\,\cite{Grazzini:2017mhc}.
This is not a trivial task due to the large numerical uncertainties associated to the $q_T^{\rm cut}\to 0$ limit.

We now discuss the method which enable us to consistently remove the FPC within the $q_T$ subtraction
formalism thus leaving
a quadratic residual uncertainty $\mathcal{O}(({q_T^{\rm cut}}/M)^2$).
The starting point is the observation that the FPC have a kinematical origin\,\cite{Ebert:2019zkb,Ebert:2020dfc,Alekhin:2021xcu}.
They are generated by the selection cuts on the final state particles and they are indeed absent in
fixed-order\,\cite{Ebert:2019zkb,Cieri:2019tfv,Ebert:2018gsn,Oleari:2020wvt} or $q_T$ resummed calculations
inclusive over the final states\,\cite{Balitsky:2017gis}
and also in the case of $q_T$ resummation with fiducial cuts when the $q_T$ recoil due to multi-parton emission is correctly taken into account\,\cite{Ebert:2020dfc}.
According to the $q_T$ resummation formalism of Refs.\,\cite{Catani:2000vq,Bozzi:2005wk}
the fiducial cross section in Eq.\,(\ref{qTsub2}) can be schematically
written in the following form:
\beqn
\widetilde\sigma_{\rm fid}^{F} =\int_{\rm cuts} d\widetilde\sigma^{F}_{LO}\otimes \mathcal{H}^F \times S(q_T,M)
+\int_{{\rm cuts}} \left[d\sigma^{\rm F+jets}_{q_T>q_T^{\rm cut}}-d\widetilde\sigma^{\rm CT}_{ q_T>q_T^{\rm cut}}\right]+\mathcal{O}\left((q_T^{\rm cut}/M)^2\right)\,,
\label{qTres}
\eeqn
where
\beqn
S(q_T,M)=\int_0^\infty db \frac{b}2 J_0(b q_T)
\exp(\mathcal{G}(\alpha_S))
\eeqn
and
\beqn
d\widetilde\sigma^{{\rm CT}} = d\widetilde\sigma^{F}_{LO} \otimes \Sigma^{F}(q_T/M) d^2\bf{q_T}\,.
\label{qTCTtilde}
\eeqn

The first term on the r.h.s.\ of Eq.(\ref{qTCTtilde}) is the resummed component of the cross section
which collects in the generalized form factor $\exp{(\mathcal{G}(\alpha_S))}$ and resums to all orders
(in the Fourier--Bessel conjugated impact-parameter $b$ space) the enhanced logarithmic corrections
of the type
$\alpha_S^n\ln^m (M^2/q_T^2)$
which are present in the
transverse momentum distribution at small $q_T$\,\cite{Bozzi:2005wk}.
The second term on the r.h.s.\ of Eq.(\ref{qTCTtilde}) is the fixed-order finite component of the cross section
and  $q_T^{\rm cut}$ represents the minimum value of $q_T$ used to compute such term. 
In the resummed formula in Eq.(\ref{qTres}) the underlying amplitude
of the Born level cross-section $d{\widetilde\sigma}^F_{LO}$, which enters also in the term $d\widetilde\sigma^{\rm {CT}}$,
differs from the corresponding
quantity $d{\hat\sigma}^{(0)}$ in Eqs.(\ref{qTsub},\ref{qTCT}) for the fact that it 
is not evaluated with the  leading order (LO) kinematics 
but
following the prescription
introduced in Appendix A of Ref.\,\cite{Catani:2015vma}, which takes into account the $q_T$ recoil originated
from the multiple radiation of soft and collinear partons in a kinematically consistent way.

Exploiting the resummation formula in Eq.(\ref{qTres}) we are thus able to construct the following modified $q_T$ subtraction formula  which is free
from linear fiducial power corrections:
\beqn
\sigma^F_{\rm fid} =\int_{\rm cuts} d\sigma^{F}_{LO}\otimes \mathcal{H}^F 
+\int_{{\rm cuts}} \left[d\sigma^{\rm F+jets}_{q_T>q_T^{\rm cut}}-d\sigma^{\rm CT}_{ q_T>q_T^{\rm cut}}\right]
+ \int_{{\rm cuts}} d\sigma^{\rm FPC}
+\mathcal{O}\left((q_T^{\rm cut}/M)^2\right)\,,
\label{qTsub3}
\eeqn
where
\beqn
 d\sigma^{\rm FPC}=\left[d\widetilde\sigma^{{CT}}_{q_T<q_T^{\rm cut}}-d\sigma^{\rm CT}_{ q_T<q_T^{\rm cut}}\right]\,.
\label{FPC}
\eeqn

The inclusion
of the additional term $d\sigma^{\rm FPC}$ for $q_T<q_T^{\rm cut}$ allows us to produce the correct behavior of the fiducial cross section
up to quadratic power corrections in $q_T^{\rm cut}$.
The terms $d\sigma^{{CT}}$ and $d\widetilde\sigma^{{CT}}$  differ for the fact that they are respectively evaluated
with the LO ($q_T = 0$) and with the recoil  ($q_T \neq 0$) kinematics. 
We note that the term $d\sigma^{\rm FPC}$ 
is universal (i.e.\ process independent)
and
it is IR finite (albeit the two terms on the r.h.s.\ of
Eq.\,(\ref{FPC}) are separately divergent).
Furthermore the contribution of $d\sigma^{\rm FPC}$
can be treated as a {\itshape local} subtraction: the
difference of the terms in Eq.\,(\ref{FPC}) is
evaluated pointwise at integrand level
and therefore the integration for $q_T<q_T^{\rm cut}$  can be extended up to 
(virtually) arbitrary small value of $q_T$. In the current numerical
implementation\,\cite{Ablinger:2018sat} we extended the $q_T$ integration of the  term
$d\sigma^{\rm FPC}$ down to $q_T/M \sim 10^{-6}$~GeV (this value is comparable with
the typical technical cuts used in local subtraction methods).

We have encoded the formula in Eq.\,(\ref{qTsub3}), by using the recoil prescription of Ref.\,\cite{Catani:2015vma}\,\footnote{In particular within the class
  of $q_T$-recoil prescriptions introduced in Ref.\,\cite{Catani:2015vma} we use the choice defined by setting the
transverse momentum of the colliding partons equal to $q_{T}/2$.},
in the public numerical program {\ttfamily DYTurbo}\,\cite{Camarda:2019zyx} which implements the $q_T$ subtraction
formalism for Drell--Yan processes.
We are thus able to confirm also numerically, up to the N$^3$LO,
that our method correctly removes the linear power corrections and to quantify the residual systematic uncertainty of the $q_T^{\rm cut}$ technical parameter.

\begin{figure}[t]
\begin{center}
\begin{tabular}{cc}
\subfigure[]{\label{fig1a}
\includegraphics[width=0.47\textwidth]{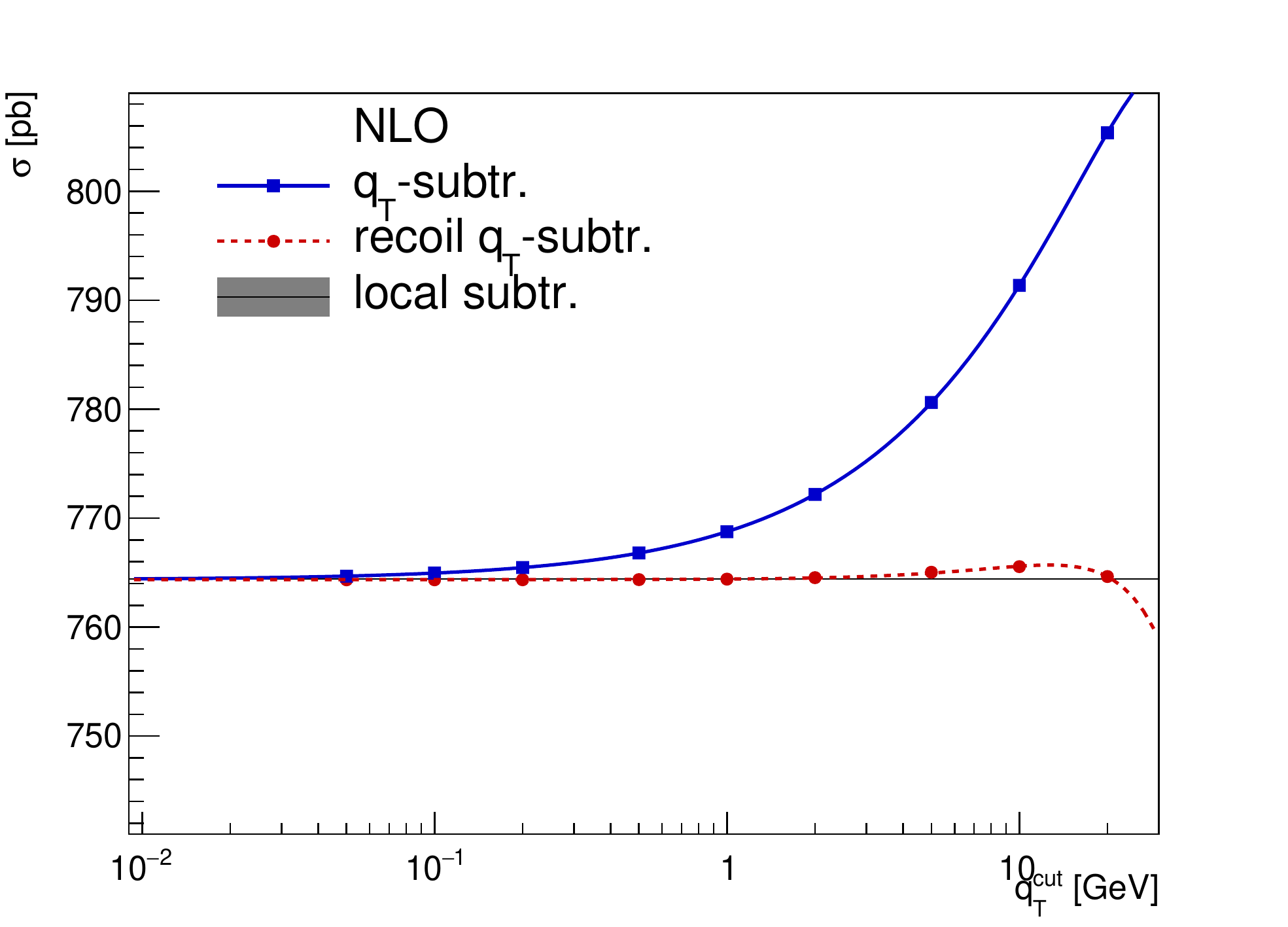}}
\subfigure[]{\label{fig1b}
\includegraphics[width=0.47\textwidth]{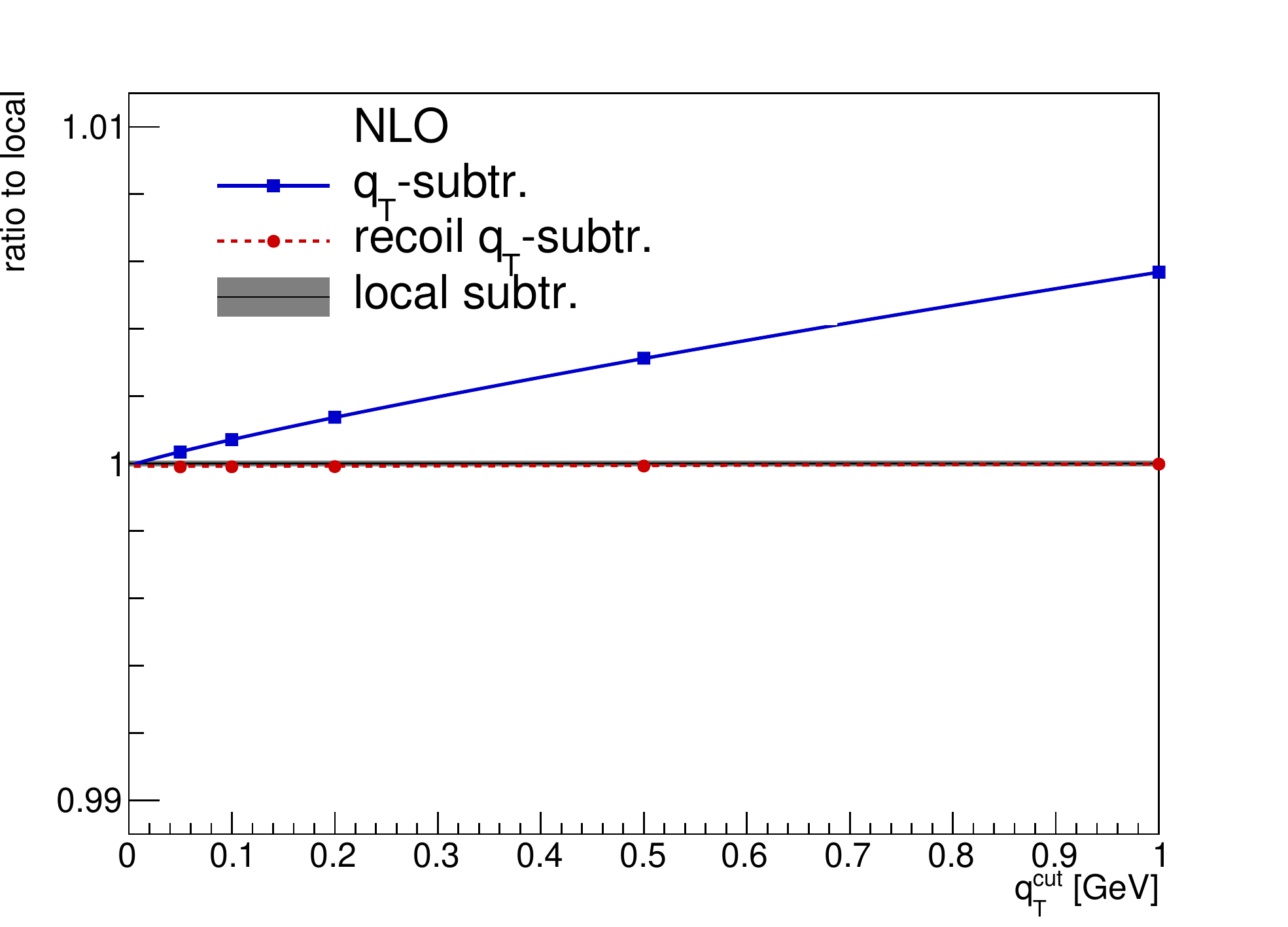}}
\end{tabular}
\end{center}
\caption{
\label{fig1}
  {\em
    Fiducial cross section for  the production of
$l^+ l^-$ pairs from  $Z/\gamma^*$ decay at
    the LHC ($\sqrt{s}=13\,$TeV).
    NLO results with the
    $q_T$ subtraction method (blue squared points) and
    $q_T$ subtraction method without FPC (red circled points)
    at various values of $q_T^{\rm cut}$, and with a local
    subtraction method (black line). Error bars indicate 
    the statistical uncertainties from Monte Carlo numerical integration.
}}
\end{figure}

We consider the production of
$l^+ l^-$ pairs from  $Z/\gamma^*$ decay at
the LHC ($\sqrt{s}=13\,$TeV) with the following
fiducial cuts\,\cite{Bizon:2019zgf}: 
the leptons are required to have transverse momentum $p_T>25\,$GeV,
pseudo-rapidity $|\eta|<2.5$ while the lepton pair system is required
to have  invariant mass $66<M<116\,$GeV and transverse momentum $q_T < 100\,$GeV.
We use parton densities functions (PDFs) from the NNPDF3.1 set\,\cite{NNPDF:2017mvq} at
NNLO  with $\as(m_Z^2)=0.118$, and we have evaluated $\as(\mu_R^2)$ at $(n\!+\!1)$-loop order
at N$^n$LO accuracy.  Factorization and renormalization scales have been set to $\mu_F=\mu_R= \sqrt{M^2+q_T^2}$.
We use the so called $G_\mu$ scheme for EW couplings
with input parameters  $G_F = 1.1663787\times 10^{-5}$~GeV$^{-2}$,
$m_Z = 91.1876$~GeV, $\Gamma_Z=2.4952$~GeV, $m_W = 80.379$~GeV\,\cite{Bizon:2019zgf}.
We then computed the fiducial cross section for the Drell-Yan process at the LHC with the
original $q_T$ subtraction formula Eq.\,(\ref{qTsub}) and with the improved formula Eq.\,(\ref{qTsub3}).

In Fig.\,(\ref{fig1}) we show the NLO fiducial
cross section  calculated for different values of the $q_T^{\rm cut}$ technical parameter 
with the original $q_T$ subtraction method (blue squared points)
and with the modified formula in Eq.\,(\ref{qTsub3}) (labeled as recoil $q_T$ subtraction, red circled points).
As a reference, we also show the result obtained with a local subtraction formalism (black line) which represents the exact
(free from significant systematic uncertainties) prediction.  The local result is obtained independently with the dipole subtraction formalism\,\cite{Catani:1996jh,Catani:1996vz} as implemented
in {\ttfamily MCFM}\,\cite{mcfm}.
Error bars in Fig.\,(\ref{fig1}) indicate 
    the statistical uncertainties from Monte Carlo numerical integration which turns out to be completely negligible.
From Fig.\,(\ref{fig1}) we observe that the systematic uncertainty (defined as the deviation from the local subtraction result)  of the original
$q_T$ subtraction results increase linearly  with $q_T^{\rm cut}$ and it is around
$0.3\%$ at $q_T^{\rm cut}=0.5\,$GeV, $0.6\%$ at $q_T^{\rm cut}=1\,$GeV, $1\%$ at $q_T^{\rm cut}=2\,$GeV and $2\%$ at $q_T^{\rm cut}=4\,$GeV.
In order to obtain a systematic uncertainty below $0.1\%$ level
a calculation with $q_T^{\rm cut}\lesssim 0.1\,$GeV is necessary. Conversely the results obtained with the $q_T$ subtraction without FPC have a systematic uncertainty
for $q_T^{\rm cut}=1\,$GeV which is smaller than the statistical uncertainty of the local-subtraction result, which is $0.01\%$.
In Fig.\,(\ref{fig1}) an interpolation of the $q_T^{\rm cut}$ dependence of the modified (original) $q_T$-subtraction obtained with quadratic
(linear and quadratic) terms is represented by the red dashed (blue solid) line.

\begin{figure}[t]
\begin{center}
\begin{tabular}{cc}
\subfigure[]{\label{fig2a}
\includegraphics[width=0.47\textwidth]{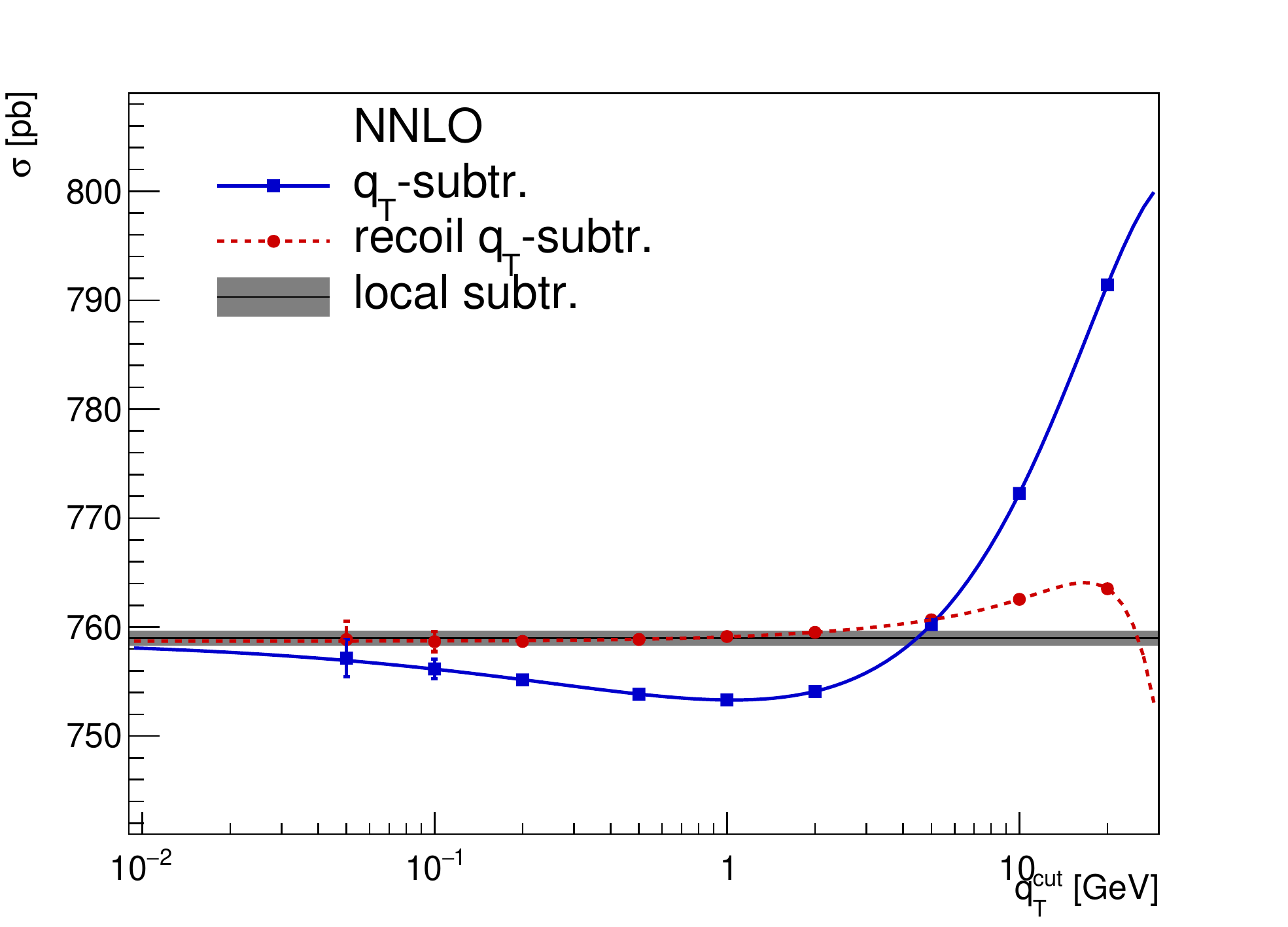}}
\subfigure[]{\label{fig2b}
\includegraphics[width=0.47\textwidth]{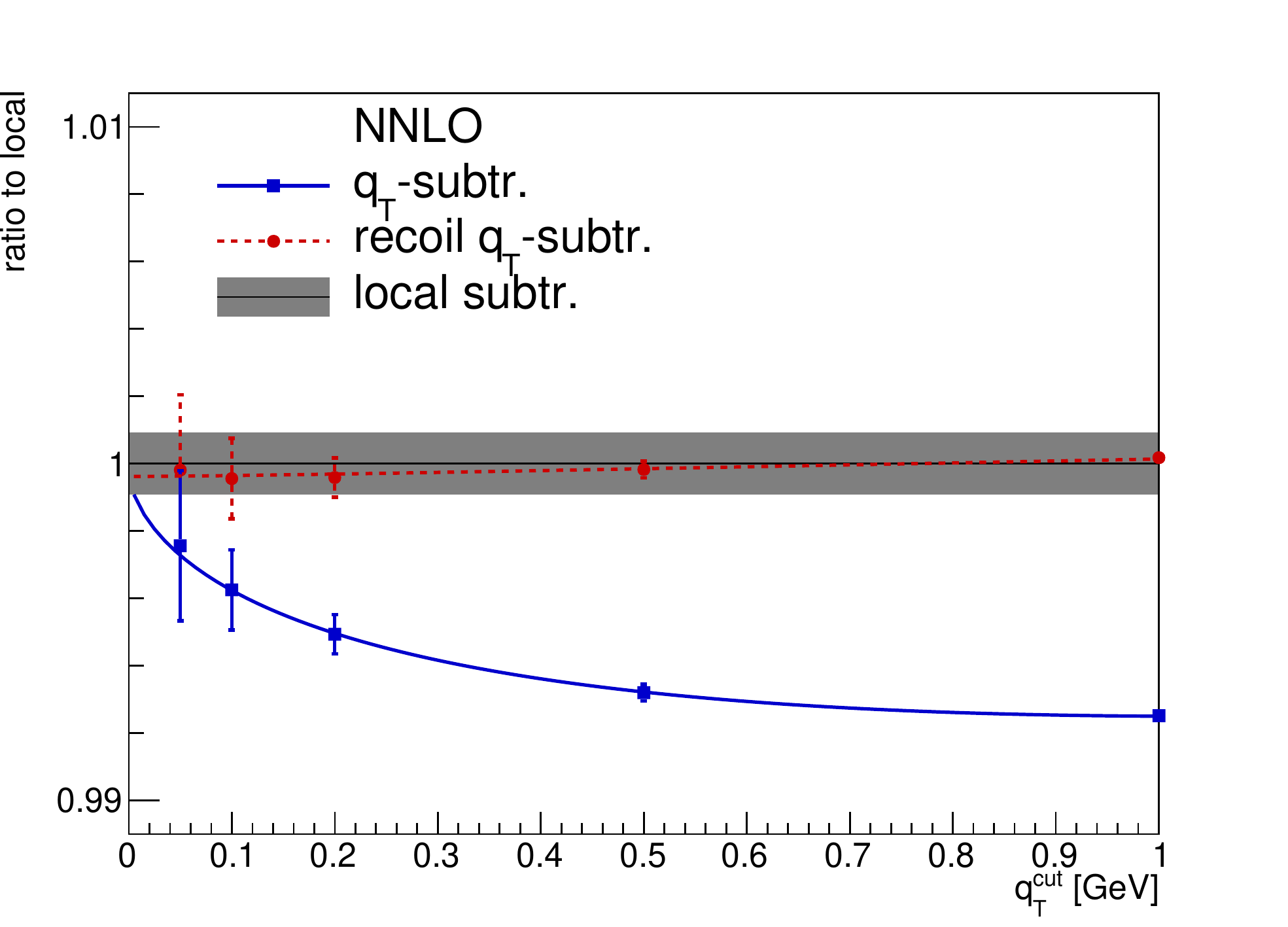}}
\end{tabular}
\end{center}
\caption{
\label{fig2}
  {\em
    Fiducial cross section for  the production of
$l^+ l^-$ pairs from  $Z/\gamma^*$ decay at
    the LHC ($\sqrt{s}=13\,$TeV).
    NLO results with the
    $q_T$ subtraction method (blue squared points) and
    the $q_T$ subtraction without FPC (red circled points)
    at various values of $q_T^{\rm cut}$, and with a local
    subtraction method (black line). Error bars indicate 
    the statistical uncertainties from Monte Carlo numerical integration.
}}
\end{figure}

In Fig.\,(\ref{fig2}) we show the fiducial
cross section at NNLO 
with the original $q_T$ subtraction method (blue squared points)
and with the modified formula in Eq.\,(\ref{qTsub3}) (red circled points)
together with the 
result obtained with a local subtraction formalism (black line).
The local result is obtained with the sector improved
subtraction formalism\,\cite{Anastasiou:2003ds,Anastasiou:2003gr} as implemented in {\ttfamily FEWZ}\,\cite{Melnikov:2006kv,Li:2012wna}.
Error bars in Fig.\,(\ref{fig2}) indicate 
the statistical uncertainties from Monte Carlo numerical integration. Statistical uncertainties 
are at the level of $0.1\%$ for the local subtraction results and 
at the level of $0.1\%$ or larger (smaller)  for the $q_T$ subtraction results with $q_T^{\rm cut}\lesssim 0.1\,$GeV ($q_T^{\rm cut}\gtrsim 0.1\,$GeV).
The $q_T^{\rm cut}$ systematic uncertainty  of the
$q_T$ subtraction results is around
$0.3\%$ at $q_T^{\rm cut}=0.5\,$GeV, $0.6\%$ at $q_T^{\rm cut}=1\,$GeV and $0.7\%$ at $q_T^{\rm cut}=2\,$GeV and $0.2\%$ at $q_T^{\rm cut}=4\,$GeV.
As in the case of the NLO results, in order to obtain a systematic uncertainty below $0.1\%$ level
a calculation with $q_T^{\rm cut}\lesssim 0.1\,$GeV is necessary. However this is exactly the IR region where  large cancellations
give rise to sizable statistical uncertainties due to numerical integration.
Conversely the results obtained with the  $q_T$ subtraction without FPC have a systematic uncertainty
which is smaller than $0.04\%$ for $q_T^{\rm cut}=1\,$GeV.
As in Fig.\,(\ref{fig1})
also in Fig.\,(\ref{fig2}) we have shown  an interpolation of the $q_T^{\rm cut}$ dependence of the results.

\begin{figure}[t]
\begin{center}
\includegraphics[width=0.7\textwidth]{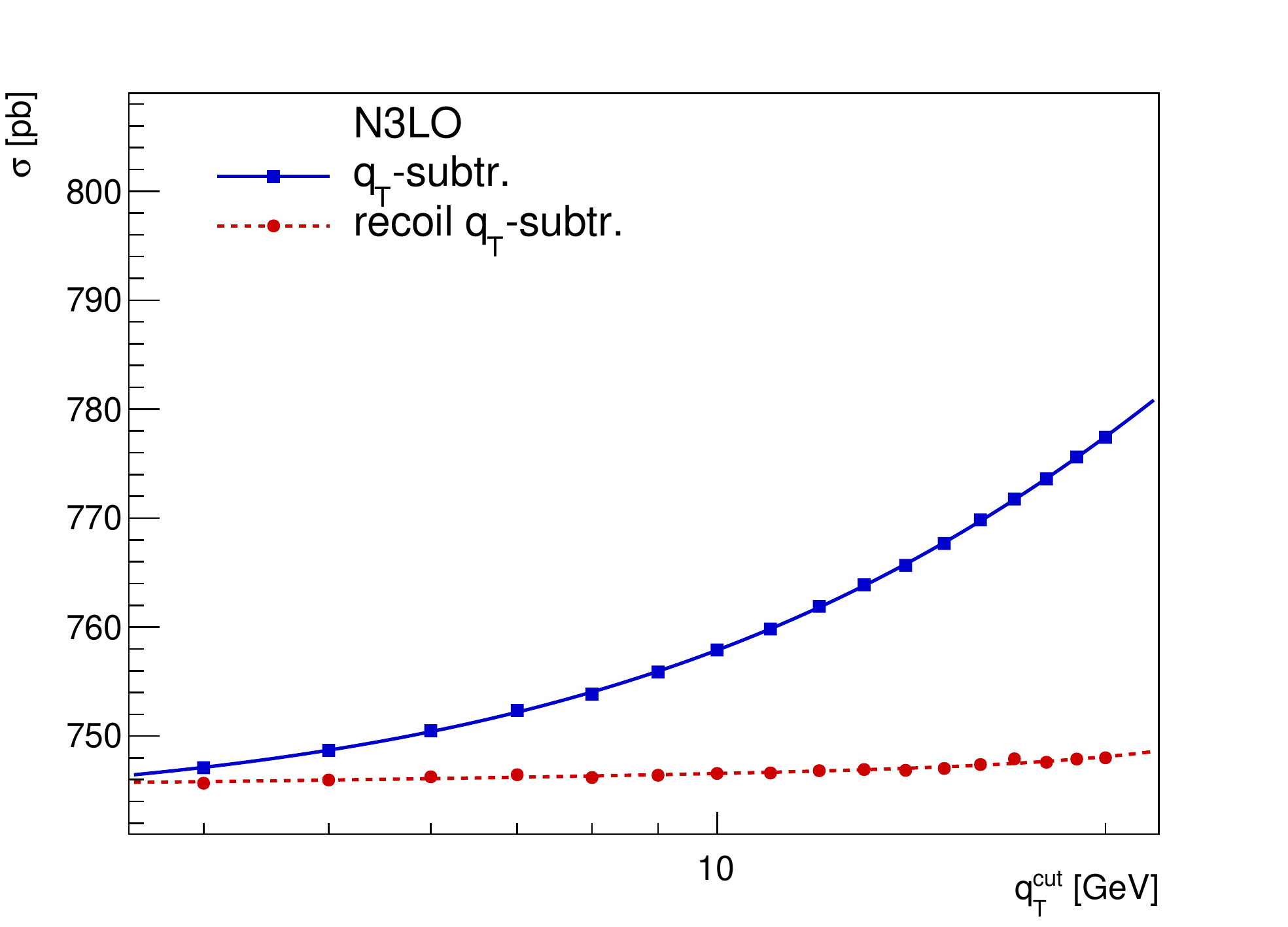}
\end{center}
\caption{
\label{fig3}
  {\em
    Fiducial cross section for  the production of
$l^+ l^-$ pairs from  $Z/\gamma^*$ decay at
    the LHC ($\sqrt{s}=13\,$TeV).
    NLO results with the
    $q_T$ subtraction method (blue squared points) and
     the $q_T$ subtraction without FPC  (red circled points)
    at various values of $q_T^{\rm cut}$. 
}}
\end{figure}

Finally, in Fig.\,(\ref{fig3}) we show the fiducial
cross section at N$^3$LO 
with the original $q_T$ subtraction method (blue squared points)
and with the modified formula in Eq.\,(\ref{qTsub3}) (red circled points)
for different values of $q_T^{\rm cut}$ with the interpolation of the results as in Figs.\,(\ref{fig1},\ref{fig2}).
No local subtraction results are available at this perturbative order.
Moreover in this case we are not able to show results for $q_T^{\rm cut}< 4$~GeV. In fact we 
have checked that our analytic expression for the counter-term $d\widetilde\sigma^{\rm CT}$ agrees  with the small-$q_T$ limit of the NNLO fixed-order results
for the production of a $Z/\gamma^*$ boson in association with jets reported in Ref.\,\cite{Bizon:2019zgf}
 at permille level down to
 $q_T\sim 4$~GeV while below that threshold such agreement
   deteriorates.
   We observe, in the case of the  $q_T$ subtraction without FPC,  a reduction of the dependence from $q_T^{\rm cut}$ which we estimate of the order of $0.1\%$ 
for $q_T^{\rm cut}\sim 4\,$GeV.

\begin{figure}[t]
\begin{center}
\includegraphics[width=0.75\textwidth]{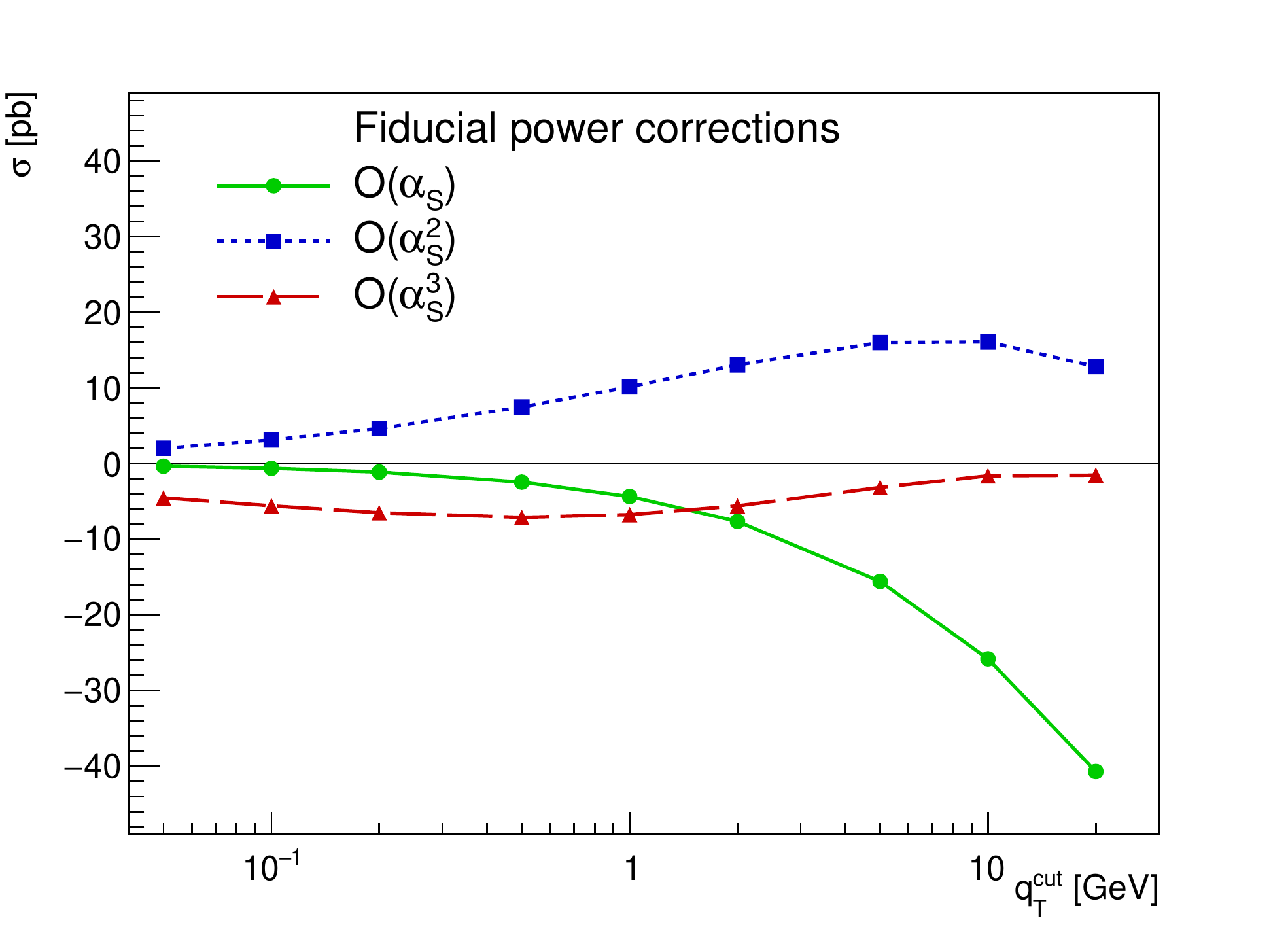}
\end{center}
\caption{
\label{fig4}
  {\em
    Production of
$l^+ l^-$ pairs from  $Z/\gamma^*$ decay at
    the LHC ($\sqrt{s}=13\,$TeV).
    Power correction contributions at $\mathcal{O}(\alpha_S)$, $\mathcal{O}(\alpha_S^2)$ and $\mathcal{O}(\alpha_S^3)$ at various values of $q_T^{\rm cut}$
}}
\end{figure}

In order to quantify the impact of the calculated fiducial power corrections, we show  in Fig.\,(\ref{fig4}) the contribution of the
FPC (Eq.\,(\ref{qTsub3})) as a function of $q_T^{\rm cut}$. First of all we observe that the {\itshape sign} of the FPC contribution changes from $\mathcal{O}(\alpha_S)$
to $\mathcal{O}(\alpha_S^2)$ and
from $\mathcal{O}(\alpha_S^2)$  to $\mathcal{O}(\alpha_S^3)$. This behaviour is consistent with the observation that linear power corrections in the small $q_T$ region
(produced by the fiducial cuts) 
results in an alternating-sign factorial growth of the fixed-order perturbative series\,\cite{Salam:2021tbm}.
The second observation is that the impact of the FPC is not numerically reduced at higher orders and it turn out to be particularly sizable at N$^3$LO  up to very small value of $q_T^{\rm cut}$:
for $q_T^{\rm cut}= 0.05\,$GeV the impact of the N$^3$LO FPC  is about $-0.4\%$ and it is the result of a $+0.3\%$ contribution 
at $\mathcal{O}(\alpha_S^2)$ and a  $-0.7\%$ at $\mathcal{O}(\alpha_S^3)$ (the $\mathcal{O}(\alpha_S)$ FPC contribution turns out to be negligible at $q_T^{\rm cut}= 0.05\,$GeV).
This means that when standard selection cuts are implemented within the original $q_T$ subtraction,  a permille level systematic accuracy
for NNLO and N$^3$LO fiducial cross sections cannot be easily reached even with extremely low values of $q_T^{\rm cut}$. 

\begin{table}[h]
  \vspace*{.5cm}
  \begin{center}
        \begin{tabular}{lcc}
      \toprule
\midrule                                                                                                                         
Order            					&  NLO   	& NNLO       	\\
\midrule                                                                                                                         
 $q_T$ subtr. ($q_T^{\rm cut}= 1\,$GeV)&  $768.8 \pm 0.1 $\,pb &  $753.3 \pm  0.3 $\,pb   \\
\midrule                                                                                                                         
 $q_T$ subtr. ($q_T^{\rm cut}= 0.5\,$GeV)&  $766.8 \pm 0.1 $\,pb &  $753.8 \pm  0.2 $\,pb   \\
\midrule                                                                                                                         
recoil $q_T$ subtr. &  $764.4 \pm 0.1 $\,pb  &  $759.1 \pm  0.3 $\,pb   \\
\midrule                                                                                                                         
local subtraction&  $764.4 \pm  0.1$\,pb &  $759.0 \pm 0.7 $\,pb \\
\midrule                                                                                                                         
\bottomrule
    \end{tabular}
  \end{center}
\caption{\em Fiducial cross sections at the LHC ($\sqrt{s}=13$~TeV): fixed-order
  results at NLO and  NNLO. 
  The uncertainties
    on the values of the cross sections refer to an estimate of the numerical uncertainties.
}
\label{table1}
\end{table}

In Table\,\ref{table1} we
report the predictions for the cross section in the fiducial region at NLO and NNLO 
with the $q_T$ subtraction method for $q_T^{\rm cut}= 0.5\,$GeV and $q_T^{\rm cut}= 1\,$GeV, 
with the recoil $q_T$ subtraction for $q_T^{\rm cut}= 1\,$GeV
and we compare  with the local subtraction results\,\footnote{Since the numerical code
  {\ttfamily FEWZ} does not allow to set $\mu_F=\mu_R= \sqrt{M^2+q_T^2}$
  the NNLO local result has been obtained with $\mu_F=\mu_R= M$. We have estimated the effect of the different scales with {\ttfamily DYTurbo}
and it turn out to be at the level of $0.5$\,pb. This effect has been included in the numerical uncertainty.}. Errors
indicate the statistical uncertainties from Monte Carlo numerical integration.
In the case of the recoil $q_T$ subtraction the results are nearly independent by $q_T^{\rm cut}$ for $q_T^{\rm cut}\lesssim \mathcal{O}$(GeV)) 

From the results of Table\,\ref{table1} 
we observe that the differences between the recoil $q_T$ subtraction  results and the
local subtraction results are of $\mathcal{O}(0.01\%)$. 
Thus the modified $q_T$ subtraction formula in Eq.\,(\ref{qTsub3}) allows us to obtain accurate permille level predictions for fiducial cross section
with values of $q_T^{\rm cut}\sim\mathcal{O}$(GeV).

\begin{table}[h]
  \vspace*{.5cm}
  \begin{center}
        \begin{tabular}{lc}
      \toprule
\midrule                                                                                                                         
Order            					&  N$^3$LO   	 \\
\midrule                                                                                                                         
 $q_T$ subtr. ($q_T^{\rm cut}= 4\,$GeV) & $747.1 \pm 0.7 $\,pb  \\
\midrule                                                                                                                         
recoil $q_T$ subtr. &  $745.7 \pm  0.7$\,pb  \\
\midrule                                                                                                                         
\bottomrule
    \end{tabular}
  \end{center}
\caption{\em Fiducial cross sections at the LHC ($\sqrt{s}=13$~TeV): fixed-order
  results at N$^3$LO. 
  The uncertainties
    on the values of the cross sections refer to an estimate of the numerical uncertainties.
}
\label{table2}
\end{table}

In Table\,\ref{table2} we
report the predictions for the cross section in the fiducial region at N$^3$LO  
with the $q_T$ subtraction method  
and with the recoil $q_T$ subtraction for $q_T^{\rm cut}= 4\,$GeV. Local subtraction results are not available at this order.

From the results shown in Fig.\,(\ref{fig4}) we could expect that a value of $q_T^{\rm cut}\sim 4$~GeV is associated  with a systematic
   uncertainty due to the FPC of around $2\%$, which is of the same order of the size of the $\alpha_S^3$ corrections and thus challenge the reliability of the $q_T$ subtraction
   results\,\footnote{To reduce such uncertainty at the few permille level in Ref.\,\cite{Camarda:2021ict} the value of $q_T^{\rm cut}\sim 4$~GeV have been used for the
     $\alpha_S^3$ terms only with a lower value of $q_T^{\rm cut}\sim 0.5$~GeV for the  $\alpha_S$ and  $\alpha_S^2$ contributions.}. However the cancellation
   of the alternating sign linear fiducial power corrections shown Fig.\,(\ref{fig4}) makes the impact of the N$^3$LO FPC for the particular value of $q_T^{\rm cut}= 4$~GeV
   to be around $0.2\%$ which is indeed the difference between the $q_T$ subtraction  and recoil $q_T$ subtraction results reported in Table\,\ref{table2}.

We finally note that any numerical implementation of the subtraction method, including the local versions, contains and depends on various technical parameters necessary to avoid the
numerical evaluation of singular points. These parameters cannot be arbitrarily large and their actual value has to be chosen in order to make the numerical result independent
(within the required numerical accuracy) from their actual value. From this viewpoint the independence by $q_T^{\rm cut}$ observed within our method 
is similar to the one observed in the local version of the subtraction method.

In this paper we have considered higher-order QCD corrections to the production of
  high-mass systems in hadron collisions within the $q_T$ subtraction formalism.
  We have presented a method to consistently remove the linear power corrections in $q_T$
  of the type $\mathcal{O}(q_T^{\rm cut}/M)$, where $q_T^{\rm cut}$ is the
  technical parameter necessary to separate resolved and unresolved
    parton emission regions,
  which appears when fiducial
  kinematical  cuts 
  are applied on the final state system. 
  As a first application we
have implemented our method within the  {\ttfamily DYTurbo} numerical program
and we have 
  considered explicitly the case of fiducial cross sections for
  $Z/\gamma^*$ boson production   at the LHC
  up to N$^3$LO
  in QCD.
We have obtained perturbative predictions
which are in excellent (permille level) agreement with those obtained with local subtraction formalism at NLO and NNLO
and we have computed  N$^3$LO predictions with a residual $q_T^{\rm cut}$ systematic uncertainty  at the permille level.

Our results can be helpful in increasing the numerical precision of the existing numerical codes based on the $q_T$ subtraction formalism and also on improving
their time performances. In particular we were able to remove the source of systematic uncertainty
at the origin of the discrepancies observed in Ref.\,\cite{Alekhin:2021xcu}.
Moreover our method is particularly important
in the cases where fully local perturbative calculations for cross section are not available
or when large numerical integration uncertainties are associated to the $q_T\to 0$ limit such
as in the case of N$^3$LO predictions in hadron collisions.

Finally, we make some observations about some consequences of our findings on resummed calculations.
Our results  show that resummed fiducial cross sections calculated by correctly taking into
account the $q_T$ recoil effects\,\cite{Catani:2015vma} (e.g.\ the resummed cross sections calculated in Ref.\,\cite{Camarda:2021ict})
are free from significant numerical systematic uncertainties due to the minimum value of $q_T$ used to compute the
finite component of the cross section in  Eq.(\ref{qTCTtilde}).
Moreover our results show that the matching between the resummed and finite (fixed-order) calculations in the small $q_T$ region has an impact of $\mathcal{O}((q_T/M)^2)$ and
it is expected to have a very small (negligible) effect for $q_T/M\lesssim \mathcal{O}(10^{-1})$ ($q_T/M\lesssim \mathcal{O}(10^{-2})$).

\paragraph{Acknowledgments.} 
We gratefully acknowledge Stefano Catani and Massimiliano Grazzini
for useful discussions  and comments on the manuscript and Alessandro Guida and Simone Amoroso for extensive tests of the numerical code.
LC is supported by the Generalitat Valenciana (Spain) through the plan GenT program (CIDEGENT/2020/011) and his work is supported by the Spanish Government (Agencia Estatal de Investigaci\'on) and ERDF funds from European Commission (Grant No. PID2020-114473GB-I00 funded by MCIN/AEI/10.13039/ 501100011033).



\begin{thebibliography}{99}

  \bibitem{Catani:1996jh}
S.~Catani and M.~H.~Seymour,
Phys. Lett. B \textbf{378} (1996), 287-301
doi:10.1016/0370-2693(96)00425-X
[arXiv:hep-ph/9602277 [hep-ph]].

\bibitem{Catani:1996vz}
S.~Catani and M.~H.~Seymour,
Nucl. Phys. B \textbf{485} (1997), 291-419
[erratum: Nucl. Phys. B \textbf{510} (1998), 503-504]
doi:10.1016/S0550-3213(96)00589-5
[arXiv:hep-ph/9605323 [hep-ph]].


\bibitem{Frixione:1995ms}
S.~Frixione, Z.~Kunszt and A.~Signer,
Nucl. Phys. B \textbf{467} (1996), 399-442
doi:10.1016/0550-3213(96)00110-1
[arXiv:hep-ph/9512328 [hep-ph]].


\bibitem{Frixione:1997np}
S.~Frixione,
Nucl. Phys. B \textbf{507} (1997), 295-314
doi:10.1016/S0550-3213(97)00574-9
[arXiv:hep-ph/9706545 [hep-ph]].

\bibitem{Catani:2007vq}
S.~Catani and M.~Grazzini,
Phys. Rev. Lett. \textbf{98} (2007), 222002
doi:10.1103/PhysRevLett.98.222002
[arXiv:hep-ph/0703012 [hep-ph]].

\bibitem{Grazzini:2017mhc}
M.~Grazzini, S.~Kallweit and M.~Wiesemann,
Eur. Phys. J. C \textbf{78} (2018) no.7, 537
doi:10.1140/epjc/s10052-018-5771-7
[arXiv:1711.06631 [hep-ph]].



\bibitem{Cieri:2018oms}
L.~Cieri, X.~Chen, T.~Gehrmann, E.~W.~N.~Glover and A.~Huss,
JHEP \textbf{02} (2019), 096
doi:10.1007/JHEP02(2019)096
[arXiv:1807.11501 [hep-ph]].

\bibitem{Camarda:2021ict}
S.~Camarda, L.~Cieri and G.~Ferrera,
[arXiv:2103.04974 [hep-ph]].

\bibitem{Ebert:2019zkb}
M.~A.~Ebert and F.~J.~Tackmann,
JHEP \textbf{03} (2020), 158
doi:10.1007/JHEP03(2020)158
[arXiv:1911.08486 [hep-ph]].

\bibitem{Ebert:2020dfc}
M.~A.~Ebert, J.~K.~L.~Michel, I.~W.~Stewart and F.~J.~Tackmann,
JHEP \textbf{04} (2021), 102
doi:10.1007/JHEP04(2021)102
[arXiv:2006.11382 [hep-ph]].

\bibitem{Alekhin:2021xcu}
S.~Alekhin, A.~Kardos, S.~Moch and Z.~Tr\'ocs\'anyi,
[arXiv:2104.02400 [hep-ph]].

\bibitem{Moult:2016fqy}
I.~Moult, L.~Rothen, I.~W.~Stewart, F.~J.~Tackmann and H.~X.~Zhu,
Phys. Rev. D \textbf{95} (2017) no.7, 074023
doi:10.1103/PhysRevD.95.074023
[arXiv:1612.00450 [hep-ph]].

\bibitem{Cieri:2019tfv}
L.~Cieri, C.~Oleari and M.~Rocco,
Eur. Phys. J. C \textbf{79} (2019) no.10, 852
doi:10.1140/epjc/s10052-019-7361-8
[arXiv:1906.09044 [hep-ph]].

\bibitem{Catani:2014qha}
S.~Catani, M.~Grazzini and A.~Torre,
Nucl. Phys. B \textbf{890} (2014), 518-538
doi:10.1016/j.nuclphysb.2014.11.019
[arXiv:1408.4564 [hep-ph]].

\bibitem{Bonciani:2015sha}
R.~Bonciani, S.~Catani, M.~Grazzini, H.~Sargsyan and A.~Torre,
Eur. Phys. J. C \textbf{75} (2015) no.12, 581
doi:10.1140/epjc/s10052-015-3793-y
[arXiv:1508.03585 [hep-ph]].

\bibitem{Buonocore:2019puv}
L.~Buonocore, M.~Grazzini and F.~Tramontano,
Eur. Phys. J. C \textbf{80} (2020) no.3, 254
doi:10.1140/epjc/s10052-020-7815-z
[arXiv:1911.10166 [hep-ph]].


\bibitem{Salam:2021tbm}
G.~P.~Salam and E.~Slade,
[arXiv:2106.08329 [hep-ph]].

\bibitem{Glazov:2020gza}
A.~Glazov,
Eur. Phys. J. C \textbf{80} (2020) no.9, 875
doi:10.1140/epjc/s10052-020-08435-4
[arXiv:2001.02933 [hep-ex]].


\bibitem{Catani:2015vma}
S.~Catani, D.~de Florian, G.~Ferrera and M.~Grazzini,
JHEP \textbf{12} (2015), 047
doi:10.1007/JHEP12(2015)047
[arXiv:1507.06937 [hep-ph]].

\bibitem{Catani:2012qa}
S.~Catani, L.~Cieri, D.~de Florian, G.~Ferrera and M.~Grazzini,
Eur. Phys. J. C \textbf{72} (2012), 2195
doi:10.1140/epjc/s10052-012-2195-7
[arXiv:1209.0158 [hep-ph]].

\bibitem{Catani:2013tia}
S.~Catani, L.~Cieri, D.~de Florian, G.~Ferrera and M.~Grazzini,
Nucl. Phys. B \textbf{881} (2014), 414-443
doi:10.1016/j.nuclphysb.2014.02.011

\bibitem{Bozzi:2005wk}
G.~Bozzi, S.~Catani, D.~de Florian and M.~Grazzini,
Nucl. Phys. B \textbf{737} (2006), 73-120
doi:10.1016/j.nuclphysb.2005.12.022
[arXiv:hep-ph/0508068 [hep-ph]].

\bibitem{Catani:2000vq}
S.~Catani, D.~de Florian and M.~Grazzini,
Nucl. Phys. B \textbf{596} (2001), 299-312
doi:10.1016/S0550-3213(00)00617-9
[arXiv:hep-ph/0008184 [hep-ph]].

\bibitem{Ebert:2018gsn}
M.~A.~Ebert, I.~Moult, I.~W.~Stewart, F.~J.~Tackmann, G.~Vita and H.~X.~Zhu,
JHEP \textbf{04} (2019), 123
doi:10.1007/JHEP04(2019)123
[arXiv:1812.08189 [hep-ph]].

\bibitem{Oleari:2020wvt}
C.~Oleari and M.~Rocco,
Eur. Phys. J. C \textbf{81} (2021) no.2, 183
doi:10.1140/epjc/s10052-021-08878-3
[arXiv:2012.10538 [hep-ph]].


\bibitem{Balitsky:2017gis}
I.~Balitsky and A.~Tarasov,
JHEP \textbf{05} (2018), 150
doi:10.1007/JHEP05(2018)150
[arXiv:1712.09389 [hep-ph]].



\bibitem{Ablinger:2018sat}
J.~Ablinger, J.~Bl\"umlein, M.~Round and C.~Schneider,
Comput. Phys. Commun. \textbf{240} (2019), 189-201
doi:10.1016/j.cpc.2019.02.005
[arXiv:1809.07084 [hep-ph]].


\bibitem{Camarda:2019zyx}
S.~Camarda, M.~Boonekamp, G.~Bozzi, S.~Catani, L.~Cieri, J.~Cuth, G.~Ferrera, D.~de Florian, A.~Glazov and M.~Grazzini, \textit{et al.}
Eur. Phys. J. C \textbf{80} (2020) no.3, 251
[erratum: Eur. Phys. J. C \textbf{80} (2020) no.5, 440]
doi:10.1140/epjc/s10052-020-7757-5
[arXiv:1910.07049 [hep-ph]].

\bibitem{Bizon:2019zgf}
W.~Bizon, A.~Gehrmann-De Ridder, T.~Gehrmann, N.~Glover, A.~Huss, P.~F.~Monni, E.~Re, L.~Rottoli and D.~M.~Walker,
Eur. Phys. J. C \textbf{79} (2019) no.10, 868
doi:10.1140/epjc/s10052-019-7324-0
[arXiv:1905.05171 [hep-ph]].

\bibitem{NNPDF:2017mvq}
R.~D.~Ball \textit{et al.} [NNPDF],
Eur. Phys. J. C \textbf{77} (2017) no.10, 663
doi:10.1140/epjc/s10052-017-5199-5
[arXiv:1706.00428 [hep-ph]].




\bibitem{mcfm}
J.~M.~Campbell and R.~K.~Ellis,
Phys. Rev. D \textbf{60} (1999), 113006
doi:10.1103/PhysRevD.60.113006
[arXiv:hep-ph/9905386 [hep-ph]];
J.~M.~Campbell, R.~K.~Ellis and C.~Williams,
JHEP \textbf{07} (2011), 018
doi:10.1007/JHEP07(2011)018
[arXiv:1105.0020 [hep-ph]];
J.~M.~Campbell, R.~K.~Ellis and W.~T.~Giele,
Eur. Phys. J. C \textbf{75} (2015) no.6, 246
doi:10.1140/epjc/s10052-015-3461-2
[arXiv:1503.06182 [physics.comp-ph]].

\bibitem{Melnikov:2006kv}
K.~Melnikov and F.~Petriello,
Phys. Rev. D \textbf{74} (2006), 114017
doi:10.1103/PhysRevD.74.114017
[arXiv:hep-ph/0609070 [hep-ph]].

\bibitem{Li:2012wna}
Y.~Li and F.~Petriello,
Phys. Rev. D \textbf{86} (2012), 094034
doi:10.1103/PhysRevD.86.094034
[arXiv:1208.5967 [hep-ph]].

\bibitem{Anastasiou:2003gr}
C.~Anastasiou, K.~Melnikov and F.~Petriello,
Phys. Rev. D \textbf{69} (2004), 076010
doi:10.1103/PhysRevD.69.076010
[arXiv:hep-ph/0311311 [hep-ph]].
\bibitem{Anastasiou:2003ds}
C.~Anastasiou, L.~J.~Dixon, K.~Melnikov and F.~Petriello,
Phys. Rev. D \textbf{69} (2004), 094008
doi:10.1103/PhysRevD.69.094008
[arXiv:hep-ph/0312266 [hep-ph]].



\end{thebibliography}
\end{document}